\documentclass[11pt,twoside]{article}
\usepackage{asp2010}
\usepackage{graphicx}

\resetcounters

\begin{document}

\title{Baryon and Sunyaev-Zel'dovich effect properties of MareNostrum and MultiDark simulated clusters (MUSIC)}
\author{Federico Sembolini$^{1,2}$, Gustavo Yepes $^1$, Marco De Petris $^2$
\affil{$^1$Departamento de F\' isica Te\' orica, M\' odulo C-15, Facultad de Ciencias, Universidad Aut\' onoma de Madrid, 28049 Cantoblanco, Madrid, Spain}
\affil{$^2$Dipartimento di Fisica,  Sapienza Universit\`a di Roma, Piazzale Aldo Moro 5, 00185 Roma, Italy}}

\begin{abstract}
We report the first results of the  MUSIC project. It consists of two  
data sets of resimulates clusters  extracted from two large dark  
matter  only simulations: Marenostrum Universe and Multidark. In  
total, the MUSIC contains more than 400  clusters resimulated with  
high resolution both with radiative and non-radiative physics  
included.  Here we present the first results on the properties of the  
baryon content and the   Sunyaev Zeldovich scaling relations.
\end{abstract}

\section{MUSIC Datasets}
The first dataset is composed of 164 resimulated clusters extracted from the MareNostrum (MN) Universe. The MN Universe \citep{marenostrum_06} is a non radiative SPH cosmological simulation of a 500 Mpc {\itshape h$^{-1}$}$^3$  cubic box with WMAP1 cosmological parameters ($\Omega_{\Lambda}$= 0.7,  $\Omega_M$= 0.3, $\Omega_b$ = 0.045, {\itshape h} = 0.7, $\sigma_8$=0.9 and a slope of {\itshape n} = 1 for the initial power spectrum). It contains a total of 2 $\times$ 1024$^3$ particles equally divided between dark matter and gas.  We selected for resimulation a total of 82 relaxed clusters and 82 bullet-like clusters and using GADGET \citep{gadget},with 8 times more particles (m$_{DM}$ = 1.03$\times$10$^9$ {\itshape h$^{-1}$ m$_{\odot}$} and m$_{gas}$ = 1.82$\times$10$^8${\itshape h$^{-1}$ m$_{\odot}$}) and adding radiative physics ({\itshape i.e.} cooling, UV photoionization, star formation and galactic winds and SN thermal and kinetic feedbacks).
The second dataset is composed of the 282 more massive (most of them with a mass M $>$10$^{15}${\itshape h$^{-1}$} m$_{\odot}$ at {\itshape z} = 0) galaxy clusters extracted from the MultiDark (MD) simulation. The MD simulation is an ART dark matter only simulation performed by Anatoly Klypin at NAS Ames, containing about 8.6 billion particles in a (1 Gpc{\itshape h$^{-1}$})$^3$ volume that was performed using WMAP5 cosmological parameters ($\Omega_{\Lambda}$= 0.73,  $\Omega_M$= 0.27, $\Omega_b$ = 0.0469, {\itshape h} = 0.7, $\sigma_8$=0.82 and {\itshape n} = 0.95). We resimulated them with 8 times more particles in a region of 6 Mpc centred around each object at z=0. We added SPH gas particles in the high resolution area and run them including radiative (as in the case of the MN dataset) and non-radiative physics. The mass resolution is m$_{DM}$ = 9.01$\times$10$^8$ {\itshape h$^{-1}$ m$_{\odot}$} and m$_{gas}$ = 1.8$\times$10$^8${\itshape h$^{-1}$ m$_{\odot}$} \citep{anatolio}. In many of the resimulation areas we often found more than one cluster. Thus, the total number of resimulated clusters we finally compiled was over 500 with M $>$ 10$^{14}$ {\itshape m$_{\odot}$}. The two datasets have now joined to form the MUSIC (MareNostrum-MUltidark SImulation of galaxy Clusters) project.

\section{Baryon properties in simulated galaxy clusters}
We analyzed some of the internal properties of the clusters. We mainly focused on the baryon properties, in order to see how relevant is the effect of overcooling on the clusters simulated with radiative physics. Overcooling is an hot topic of cluster simulations \citep{krav2005}: the huge star formation rate in the cluster center implies a star fraction larger than the observed one. We studied the behavior of a set of cluster properties (total mass, gas and star mass, baryon , star and gas fraction) at six different overdensities ranging from $\Delta$ = 200 to $\Delta$ = 2500, where $\Delta$ is the ratio between cluster density and local critical density. These two limit values correspond to an overdensity ($\Delta$ = 200) at  which the cluster mass well approximates the virial mass and a value  ($\Delta$ = 2500) often referred in X-ray observations allowing us to study the effect of overcooling. We also analyzed the dataset at five different redshifts (from {\itshape z} = 0 to {\itshape z} = 0.43) in order to study the effect of cluster evolution. We define the overdensity radius as the radius of the sphere within which the average density of the cluster is $\Delta$ times the critical density:
\begin{equation}
\frac{4}{3}\pi \rho_c(z)\Delta r^3_{\Delta}=M_{TOT}(r_{\Delta})
\end{equation}
The critical density is defined as {\itshape $\rho _c$(z)=3H$_0$$^2$E(z)$^2$/8$\pi$ G} depending on the assumed cosmology.
We calculated the gas fraction as {\itshape f$_{gas}$ $\equiv$ M$_{gas}$/M$_{TOT}$}; similarly, the star and baryon fraction as {\itshape f$_{star}$ $\equiv$ M$_{star}$/M$_{TOT}$} and {\itshape f$_{bar}$ $\equiv$(M$_{star}$+M$_{gas}$)/M$_{TOT}$}. Figures 1 and 2 show the baryon, gas and star fractions along the redshift for clusters of the MD dataset simulated with radiative physics at two different overdensities. In {\itshape fig.1} $\Delta$ = 500 and the mean values agree very well with observations\citep{gas}: the baryon fraction is close to the critical ratio $\Omega _b$/$\Omega _m$ as expected and the gas fraction is about twice bigger than the star fraction. In {\itshape fig.2} $\Delta$ = 2500 the effect of overcooling  (when the clusters are less evolved). The radiative clusters of MN dataset show a similar behavior but, being the masses smaller than those of MD dataset, the effect of overcooling is slightly stronger and the star fraction higher. On the other hand, morphology seems not to affect the baryon properties. Non-radiative clusters are obviously not affected by overcooling and the mean gas fraction approaches the critical ratio as we move towards $\Delta$ = 200.

\begin{figure}[!ht]
\begin{center}
\includegraphics[width=250pt]{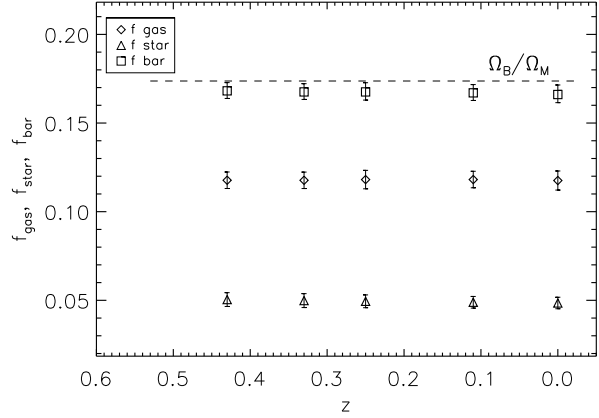}
\caption{Baryon (squares), gas (diamonds) and star (triangles) fraction {\itshape vs} redshift at overdensity $\Delta$ = 500 of MD clusters resimulated with non-adiabatic physics}
\end{center}
\end{figure}

\begin{figure}[!ht]
\begin{center}
\includegraphics[width=250pt]{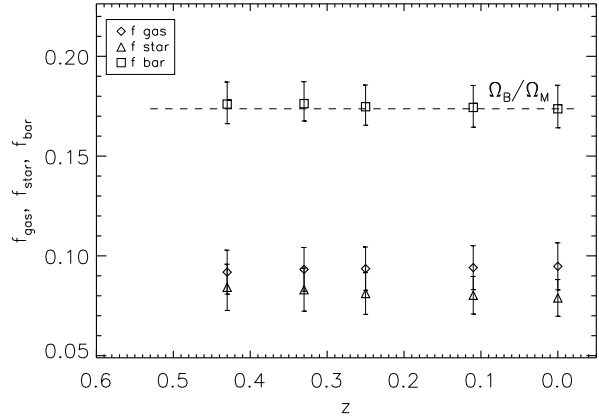}
\caption{Baryon (squares), gas (diamonds) and star (triangles) fraction {\itshape vs} redshift at overdensity $\Delta$ = 2500 of MD clusters resimulated with non-adiabatic physics}
\end{center}
\end{figure}

\section{SZ scaling relations}
The integrated brightness of the thermal component of the Sunyaev-Zel'dovich effect (SZ) \citep{SZ70}, Y, is one of the strongest proxy to estimate the mass of galaxy clusters.Y is the integration over the solid angle subtended by the cluster  $\Omega$ of the Compton-y parameter defined as the integration of the electronic pressure along the line of sight. 
 We studied the {\itshape Y-M} scaling relation of our simulated cluster sample to check it and compare it with other simulations, theory and observations.
Under the assumption of an isothermal model the integrated {\itshape Y} is proportional to the integral of the electron density {\itshape n$_e$}:
\begin{equation}
YD^2_A\propto T_e\int n_e dV=M_{gas}T_e=f_{gas}M_{TOT}T_e
\end{equation}
where {\itshape T$_e$} is the electronic temperature and {\itshape D$_A$} the angular distance
If we assume that cluster formation is driven by a spherically gravitational processes mainly due to the dark matter component with baryons and dark matter in hydrostatic equilibrium \citep{k1986} we find a scaling relation based on self-similarity \citep{bona2006}:

\begin{equation}
YD^2_A\propto f_{gas}M^{5/3}_{tot}E(z)^{2/3}
\end{equation}
where {\itshape E(z)} is the scale evolution. If we study this scaling relation in the log space, in the form {\itshape $\log$(Y) = A$\log$(X)+B} we expect a slope {\itshape A = 5/3 = 1.66}. To estimate Y for each cluster we built SZ maps by Compton-{\itshape y} parameter :
\begin{equation}
y \simeq \frac{k_B\sigma_T}{m_ec^2}\sum_in_{e,i}T_{e,i}W(\mid {\bf r}-{\bf r_{cdm}}\mid,h_i)
\end{equation}
where {\itshape h$_i$} is the smoothing length of the particle and {\itshape W} the particle kernel, {\itshape $\mid$r- r$_{cdm}\mid$} the position of SPH the particle respect to the centre of mass. All Y-M scaling relations, independently from physics, redshift or overdensity, showed a slope at least very close to the expected value of {\itshape A = 1.66} (see {\itshape fig.3})

\begin{figure}[ht!]
\begin{center}
\includegraphics[width=250pt]{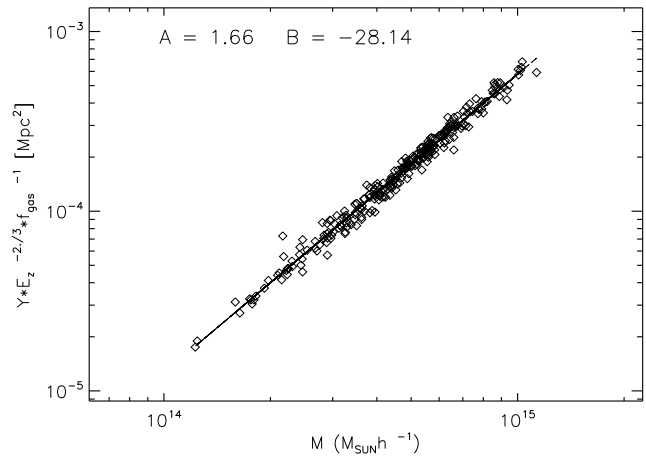}
\caption{{\itshape Y-M} scaling relation of MD radiative clusters at {\itshape z = 0}, $\Delta$ = 500}
\end{center}
\end{figure}

Concluding, we analyzed two datasets of clusters for a total of more than 600 objects simulated with high resolution and both radiative and non-radiative physics. The effect of overcooling becomes relevant only at high overdensities and the Y-M scaling relations show a very good agreement with a self-similar model. 

\bibliographystyle{asp2010}
\bibliography{author}

\begin{thebibliography}{}
\expandafter\ifx\csname natexlab\endcsname\relax\def\natexlab#1{#1}\fi
\expandafter\ifx\csname url\endcsname\relax
  \def\url#1{\texttt{#1}}\fi
\expandafter\ifx\csname urlprefix\endcsname\relax\def\urlprefix{URL }\fi
\providecommand{\eprint}[2][]{\url{#2}}

\bibitem[{{Bonamente} et~al.(2006){Bonamente}, {Joy}, {LaRoque}, {Carlstrom},
  {Reese}, \& {Dawson}}]{bona2006}
{Bonamente}, M., {Joy}, M.~K., {LaRoque}, S.~J., {Carlstrom}, J.~E., {Reese},
  E.~D., \& {Dawson}, K.~S. 2006, \apj, 647, 25.
  \eprint{arXiv:astro-ph/0512349}

\bibitem[{{Gottlober} et~al.(2006){Gottlober}, {Yepes}, {Khalatyan}, {Sevilla},
  \& {Turchaninov}}]{marenostrum_06}
{Gottlober}, S., {Yepes}, G., {Khalatyan}, A., {Sevilla}, R., \& {Turchaninov},
  V. 2006, in The Dark Side of the Universe, edited by {C.~Manoz \& G.~Yepes},
  vol. 878 of American Institute of Physics Conference Series, 3.
  \eprint{arXiv:astro-ph/0610622}

\bibitem[{{Kaiser}(1986)}]{k1986}
{Kaiser}, N. 1986, \mnras, 222, 323

\bibitem[{{Klypin} et~al.(2001){Klypin}, {Kravtsov}, {Bullock}, \&
  {Primack}}]{anatolio}
{Klypin}, A., {Kravtsov}, A.~V., {Bullock}, J.~S., \& {Primack}, J.~R. 2001,
  \apj, 554, 903. \eprint{arXiv:astro-ph/0006343}

\bibitem[{{Kravtsov} et~al.(2005){Kravtsov}, {Nagai}, \&
  {Vikhlinin}}]{krav2005}
{Kravtsov}, A.~V., {Nagai}, D., \& {Vikhlinin}, A.~A. 2005, \apj, 625, 588.
  \eprint{arXiv:astro-ph/0501227}

\bibitem[{{LaRoque} et~al.(2006){LaRoque}, {Bonamente}, {Carlstrom}, {Joy},
  {Nagai}, {Reese}, \& {Dawson}}]{gas}
{LaRoque}, S.~J., {Bonamente}, M., {Carlstrom}, J.~E., {Joy}, M.~K., {Nagai},
  D., {Reese}, E.~D., \& {Dawson}, K.~S. 2006, \apj, 652, 917.
  \eprint{arXiv:astro-ph/0604039}

\bibitem[{{Springel}(2005)}]{gadget}
{Springel}, V. 2005, \mnras, 364, 1105. \eprint{arXiv:astro-ph/0505010}

\bibitem[{{Sunyaev} \& {Zeldovich}(1970)}]{SZ70}
{Sunyaev}, R.~A., \& {Zeldovich}, Y.~B. 1970, \apss, 7, 3

\end{thebibliography}

\end{document}